  \documentclass[12pt]{article}
\title{N=4}
\usepackage{mathrsfs}
\usepackage{amsmath}

\global\arraycolsep=1pt
\usepackage[colorlinks=true,backref=true,linkcolor=black,anchorcolor=black,citecolor=black,filecolor=black,menucolor=black,pagecolor=black,urlcolor=black]{hyperref}
\usepackage[Symbol]{upgreek}
\usepackage{bbm}  
\usepackage{dsfont}
\usepackage{amssymb}
\usepackage{textcomp}

\newcommand{\scal}[1]{\bigl ({#1} \bigr )}
\def\bea{\begin{eqnarray}} 
\def\eea{\end{eqnarray}}
\def\be{\begin{equation}}
\def\ee{\end{equation}}
\newcommand{\CR}{\nonumber \\*}

\newcommand{\trace}{\hbox {Tr}~}

\def\L{{\cal L}}

\DeclareMathAlphabet{\mathpzc}{OT1}{pzc}{m}{it}

\def\c{\mathfrak{c}}

\def\A{a}

\def\F{\mathscr{F}}

\def\t{\tilde}

\usepackage[colorlinks=true,backref=true,linkcolor=black,anchorcolor=black,citecolor=black,filecolor=black,menucolor=black,pagecolor=black,urlcolor=black]{hyperref}

\def\z{\zeta}

\def\susy {\delta^{\scriptscriptstyle \,Susy}}

\usepackage[Symbol]{upgreek}
\usepackage{caption}
\usepackage{bbm}
\usepackage{dsfont}
\usepackage{amssymb}
\usepackage{textcomp}
\def\bea{\begin{eqnarray}}
\def\eea{\end{eqnarray}}
\def\be{\begin{equation}}
\def\ee{\end{equation}}

\def\L{{\cal L}}

\def\susy{{\delta^{\mathpzc{Susy}}}}

\def\t{\tilde}

\begin{document}
\allowdisplaybreaks[1]
\renewcommand{\thefootnote}{\fnsymbol{footnote}}
\def\N{\mathcal N }
\begin{titlepage}
\begin{flushright}
CERN-PH-TH-2008-191\\
\end{flushright}
\begin{center}
 {{\Large \bf
$\N=4$  Yang--Mills theory \\  as a complexification of the  $  \N=2$ theory }}
\lineskip .75em
\vskip 3em
\normalsize
{\large Laurent  Baulieu\footnote{email address: baulieu@lpthe.jussieu.fr}}
\\
 \it Theory Division, CERN,  Switzerland\footnote{
1211-Geneve 23, Switzerland}
 \\
 {\it   LPTHE, Universit\'es Pierre et Marie Curie,  Paris,
France}\footnote{
4 place Jussieu, F-75252 Paris Cedex 05, France.}
\footnote{Based on lectures given in Carg\`ese, June 16-28, 2008.}


\vskip 1 em
\end{center}
\vskip 1 em
\begin{abstract}
A   complexification of the twisted  $\N=2$ theory allows one to  determine    the  $N=4$   Yang--Mills theory in its third twist  formulation.    The imaginary part of the gauge symmetry is  used  to eliminate  two scalars fields and create   gauge covariant  longitudinal components   for   the imaginary part of the gauge field. The latter becomes the    vector field of the   thirdly twisted $\N=4$ theory. Eventually, one gets a one to one correspondence between the fields of both theories. 
  Analogous  complexifications can   be done for topological 2d-gravity and topological sigma models.
   \end{abstract}

\end{titlepage}
\renewcommand{\thefootnote}{\arabic{footnote}}
\setcounter{footnote}{0}

\def\N{{\cal{N}}}

\section{Introduction}

The $\N=4, d=4$ super-Yang--Mills theory can be expressed  under  three different   twisted versions~\cite{marcus}\cite{MQ}.     The third twisted version  was originally found by Marcus~\cite{marcus}. Kapustin and Witten have  beautifully interpreted    part of the Langland program in terms of   its quantum field theory~\cite{KW}.

   The    information about the $\N=4 $  theory in flat space  can be  encoded in any one  of its  twisted or untwisted versions,    since these theories   are   related by linear field transformations.   These   formulations give  different insights for the $\N=4$ supersymmetry.
Their links are most likely preserved by radiative corrections.    Here, 
it  will   be  shown  that the  formalism   for  the  thirdly twisted   $N=4$ supersymmetric Yang--Mills theory  is very analogous to  the apparently  much simpler one of the twisted $\N=2$ theory~\cite{DW}\cite{basi}, modulo a complexification of the latter theory,    followed by a gauge-fixing of the imaginary part of its gauge symmetry. 
    From the point of view of the Poincar\'e supersymmetry, it   comes as a surprise that such a simple link exists between the $\N=2$ and $\N=4$ cases.  This   might  suggest  the existence of   a  possibly interesting new  twisted  superfield description of the $N=4 $ theory, with a superspace with~5=4+1 anticommuting directions, as in \cite{TSP}.        

\def\t{\tilde}
The general idea is as follows.
 The twisted $\N=2$ theory   expresses  the vector supersymmetry multiplet   with eight fermionic degrees of freedom  as       
\bea\label{N=2}
A_\mu,   \Psi_\mu,    \chi_{\mu\nu^-},    \eta  , 
  \Phi ,      \bar\Phi  
  \eea
     The  corresponding   quantum field theory  and its  algebraic structure    are   now quite well understood, both in  twisted superspace and   component formalism~\cite{N=4}\cite{TSP}. An  advantage of  using  the  twisted formulation    is that it    isolates an interesting sub-sector of the   $\N=2$ supersymmetry,    made of 5 =1+4 generators,  a scalar and a vector. This fact  generalizes itself   to the case   of extended supersymmetry, and gives an interesting way of exploring its off-shell structure.

As,\ we  will     show,       the twisted $\N=2$ theory   accommodates quite well a complexification of the  fields  in~(\ref{N=2}), as follows     
\bea \label{comp}
   A_\mu+i V_\mu, \ \   \Psi_\mu +   i  \t  \Psi_\mu,   \ \  \chi_{\mu\nu^-}     +i  \chi_{\mu\nu^+},  \ \    \eta   +i  \t {\bar  c }, \ \  \Phi   +i\t \Phi, \ \   \bar\Phi +  i\t {  \bar\Phi}\eea
and   it yields  the information about the $\N=4$ theory, expressed in its third twist.

The complexified  multiplet~(\ref{comp})  is covariant under complex  infinitesimal gauge   transformations, where  the parameters have been also    complexified.  It permits one to define          a "complexified $\N=2$ supersymmetric gauge theory",  which, however,   seems quite different than     the $\N=4$~theory.  Its  field content  is made of   16 fermions as in the~$\N=4$ theory,  but there are troublesome features:  the   complex gauge invariance introduces non-unitarity questions;  there are 4 scalar fields instead of 6, and one has  an imaginary part to the gauge field. However, on-shell, the good point is that the  counting sounds right, since, by combining the complex part of the    gauge invariance and the on-shell condition,   $V_\mu$ has two degrees  of freedom, which, added to the~4 degrees of freedom carried by the fields  
$ \Phi   ,  \t \Phi,    \bar\Phi , \t {  \bar\Phi}$,   give a number of  6~on-shell   bosonic    degrees of  freedom that may represent the 6 scalar fields   of the   standard $\N=4$ theory. In fact, this gauge theory has not yet been studied in details, as one could attempt by a standard covariant gauge-fixing of the complex gauge field $A+iV$ \footnote{ Notice       that    \cite{cswitten} has already pointed out the interest of the Chern-Simons theory with a complex gauge invariance.}. Here, we will   propose in a rather unorthodox way   that one can gauge-fix the imaginary part of the gauge symmetry in a supersymmetric way,  which also preserves   the   real part of the gauge symmetry. In this way  $\bar\Phi , \t {  \bar\Phi}$ become equal to zero and $V^\mu$ becomes  ``covariantly  gauge-fixed",  by addition of   a  term $(D_A^\mu V_\mu )^2$ in the action, which gives 
   a  propagating longitudinal part that  is covariant under  the real part of the gauge symmetry. Then,       $ V^\mu$ 
becomes  a vector field that only transforms tensorially under the gauge symmetry, and 
 effectively describes 4~degrees of freedom, by mean of an action
 $V^\mu D^2_A V_\mu  +{\rm interacting  \   terms}$.
 Quite surprisingly,   the  $\N=4 $   theory is  then recovered, in    the  third twist formulation of Marcus~\cite{marcus}. Moreover,  some of  the  $\N=4$   transformation  laws can be set under a form that   might  permit   an analogous geometrical interpretation as in earlier works about  the Donaldson--Witten theory~\cite{basi}. 


This presentation is organized as follows. In a first section, there is    a heuristic derivation of the thirdly  twisted   $\N=4$ supersymmetry  and of its relationship with the twisted  $\N=2$ supersymmetry, modulo a complexification.  This indirect derivation actually provides   many of   
 the   final formula,  but it may appear as quite formal. In fact,  
it  can be skipped, and one can proceed directly to  the next sections,  where the construction  of the $\N=4$ theory  is explained in a  self-contained way, starting   from a  complexification of the $\N=2$ theory and of its horizontality equation. In the last section, there are analogous considerations   for the topological sigma-model and the topological 2d-gravity.

\section{Some remarks about the horizontality conditions for the $\N=4$ theory}

\subsection{First twist horizontality conditions}

   Maximal supersymmetry   in 8 dimensions  is a useful  theory for understanding the various twists of the $\N=4, d=4 $ theory. It  has a     twisted formulation~\cite{basi},   which   is        analogous to  that of    the $\N=2,d=4$ theory, but originates from triality,   with twisted scalar and vector generators~ \cite{TSP} \cite{shadow}.  Its dimensional reduction 
indicates  that, for the  $N=4$ supersymmetric Yang--Mills theory in  euclidean  flat space,  one has four differential operators  $s,\bar s , \delta , \bar \delta 	$, which are 		defined by graded   horizontality conditions with Bianchi identities, as follows~\cite{N=4} 
\begin{multline}\label{hcA}
(d + s + \bar s + \delta + \bar \delta) \scal{ A + c  + \bar c +
  {i_  \kappa}{ { \gamma_1 }}+ {i_  {\bar  \kappa} }\bar { \gamma_1 } } + \scal{
  A+ c + \bar c + {i_  \kappa}{ \gamma_1 } + {i_ {\bar   \kappa}}\bar { \gamma_1 } }^2 \\*
= F + \Psi + \bar\Psi + g(\kappa)  {\eta }       +    g(\bar   \kappa)    \bar \eta+ 
g(J_I   \kappa)  {\chi{  ^I }}  
+
g(J_I  \bar \kappa)       \bar \chi^{I  }   + ( 1 + \kappa      \cdot  \bar \kappa ) \scal{ \Phi +
  L + \bar \Phi }  
\end{multline}  
\vspace{-5mm}
\be\label{hch}
(d_A + s_c + \bar s_{\bar c} + \delta + \bar \delta ) h^I = d_A h^I + \bar\chi^I - \chi^I + i_{J^I \kappa}
\scal{\bar\Psi - \Psi}                  
\ee
 
\begin{multline}
(d + s + \delta - i_\kappa) \scal{ A + c + {i_  \kappa}{ \gamma_1 } } + \scal{
  A+ c + {i_  \kappa}\gamma_1}^2 \\*
= F + \Psi + g(\kappa) \eta + g(J_I \kappa) \chi^I + \Phi + |\kappa|^2
\bar \Phi  
\end{multline}
\vspace{-5mm} 
\be (d_A + s + \delta_\gamma - i_\kappa) h^I 
+[   A + c + {i_  \kappa}{ \gamma_1 }   ,  h^I   ]
= d_A h^I + \bar\chi^I +
i_{J^I \kappa} \bar\Psi 
\ee 

  The space is Euclidean. 
 $A=A_\mu dx^\mu $ is the 1-form gauge connection.  The anticommuting       fields $ \Psi_\mu,  \bar\Psi_\mu, \chi^I , \bar\chi^I ,\eta$ and $\bar\eta$ represent the four Majorana spinors of the $N=4$ theory   in the first twist formulation,  for a total of   of sixteen  (off-shell)  anticommuting   degrees of freedom. 
The  1-forms   $\Psi \equiv \Psi _\mu dx^\mu,$ and $  \bar  \Psi \equiv   \bar  \Psi_\mu dx^\mu         $ are commuting objects because for instance     the anticommuting field  $\Psi_\mu $ is  multiplied by  $dx^\mu$.   The  bosonic    scalar     fields
  $ \Phi ,
  L , \bar \Phi, h^I$ are the first twist transforms of the  6-dimensional     $R$-symmetry multiplet scalar  field  of  the $N=4$ theory  \cite{Yamron}\cite{Vafa}.  In this twist,  the $R  \times SO(4)$ symmetry has been reduced to a smaller but large enough subgroup,
   by taking the diagonal of one $SU(2)$ factor of the $R=SO(5,1)$ euclidean $R$-symmetry with one $SU(2)$ factor of the original $SO(4)$  Lorentz symmetry. This redefines a new Lorentz symmetry  $SO(4)'$ for the twisted fields, so that the  remaining global invariance is       a $SO(4)'\times  SL(2,R)   $ symmetry.  The $SL(2,R)$ triplet index $I=1,2,3$ can be identified as a self-dual index of $SO(4)'$, by mean of Kahler invariant forms.

  In the  above equations, $\kappa$ and  $\bar \kappa$  are arbitrarily given constant vectors,  each one with 4 commuting components  $\kappa^\mu $ and  $\bar \kappa  ^\mu$.  They can  considered as a set of 8 independent parameters.  Here and elsewhere $g(\kappa)$ is the one form $g(\kappa)=g_{\mu\nu }dx^\nu$ and  
  $i_\kappa$ is 
  the contraction operator  along  $\kappa$, eg,  $i_\kappa  dx^\mu =\kappa^\mu$.

       The  graded   operator       $\delta$  and $\bar \delta$ are  in correspondence with vector anticommuting differential  operators  $\delta_\mu $  and $\bar \delta_\mu$, with   $\delta     \equiv   \kappa^\mu      \delta_\mu $  and 
   $\bar  \delta     \equiv   \bar \kappa^\mu   \bar   \delta_\mu $.   The
   transformation laws of all fields  under the action of $s,\bar s , \delta , \bar \delta 	$ is  obtained by decomposition of  the above equations on all possible polynomials in  $\kappa$ and  $\bar \kappa$.

  The anticommuting   scalar  fields   $ c  , \bar c $ are called   scalar   shadows and 
  $ { \gamma_1 } $ and $ \bar \gamma _1$ are     commuting 1-form  shadows. The    scalars      $ c  , \bar c 
  $ and ${i_  \kappa}\gamma ,{i_  \kappa}\bar \gamma  $ must not be confused with Faddeev--Popov ghosts. They have very different transformation laws, and they  play  very different roles~: they are used  for keeping track of supersymmetry at the quantum level~\cite{TSP}.  In fact, 
 the introduction of shadows allows one to have   no gauge transformations     occurring  in the squared of the differentials    $s,\bar s , \delta , \bar \delta 	$.

   As shown in \cite  {N=4},     it is only for    $ \kappa =\bar  \kappa$ that   no  equation of motion occurs in the commutators of these differential operators. 
   This can be verified by 
    solving  the equations and       computing   the action of  $s,\bar s , \delta , \bar \delta 	$   on all fields.   With this restriction $ \kappa =\bar  \kappa$, the 
    above horizontality condition determine    six  generators  of the $\N=4$ theory, made of two scalars   and one  vector, under the form of differential operators.. It was shown that the invariance under this twisted supersymmetry with 6 generators unambiguously determines the $\N=4$ action.

The  equation  (\ref{hch}) for  the field $h^I$  seems however  to escape     any  kind of a geometrical interpretation, contrarily to  that satisfied by the field $A$ in (\ref{hcA}), which is a     curvature equation. 
This motivates a further change of variable for getting much better looking equations, the result being that one must  switch  from  the first twist of  \cite{Vafa}  to the third one, which  was   
  found  in   \cite{marcus} \footnote{Let us recall that, from  a  purely 4-dimensional   point of view,   a   systematic analysis of all possible way of  extracting $SU(2)$ factors from  the $R=SO(5,1)$ symmetry of the $\N=4$ theory allows one to find all  its possible independents twists.}.

 \subsection{  Heuristic    switch to the   third     twist horizontality conditions}

 
   \def\mQ{{\mathcal Q}}
   To proceed, we have to do a slight digression, coming back to the untwisted formalism. 
To make a bridge between twisted and untwisted fields,  one can   first   define the N=4~supersymmetry ``almost-differential"  operator $\mathcal Q$ for the untwisted fields,      as follows~\cite{N=4}
    \be\label{magic}
 (d+\mQ  -i   _     {    \bar  \epsilon \gamma^\mu \epsilon}     )(A+c) +(A+c)^2=F+\susy A  +  \bar  \epsilon[  \phi ]\epsilon
 \ee 
 
 Here, there is a single scalar shadow field $c$, and the  generic   parameter  $\epsilon$  of supersymmetry appears explicitly. This equation defines a    $\mQ$-transformation of fields  in function of    the   scalar shadow $c$ and the  supersymmetry commuting parameter~$\epsilon$. The later  is made of   4  Majorana spinors, that count for a total of  16 real parameters, and
$\mQ \epsilon =0$.  In fact, Eq.~(\ref{magic}) shows that the $\mQ$ transformation of a classical field can be identified as a supersymmetry transformation  $\susy(\epsilon) $  minus a compensating gauge transformation  $\delta^{\rm gauge}(c)$, with a  local anticommuting parameter equal to the shadow field.    The  four gluinos  determine  a  $SL(2,\mathds{H})$-Majorana spinor,   as well as      $\epsilon$. The six scalar fields build    a real antisymmetric representation of $SL(2,\mathds{H}) $ whose Lie algebra is isomorphic to $SO(5,1)$.
 
Thank's to the shadow dependence,      no gauge transformation occurs in the expression of $\mQ^2$.   However for a general  $\epsilon$, $\mQ^2$ closes  on translations   only    modulo some equations of motion of a $Q$-invariant lagrangian. To get rid of equations of motion in the closure relations,  one must restrict the supersymmetry parameters. This    can be done consistently when one twists the fields, since this operation  allows one to  lower the size of the symmetry. In fact,  the   twisted  representations for the supersymmetry parameters (and more generally all $SO(5,1)$-Majorana spinor-tensors)  become reducible, and can be separated in shorter multiplets. It becomes possible to consistently reduce the size of the  $\N=4$ supersymmetry,   by  projecting $\epsilon$ in these reduced representations, while retaining some of its non-trivial features.

As shown in \cite{marcus} there    are in fact   three  possible twists in the  $\N=4$  superPoincar\'e theory, due the three  possibilities of selecting a $SU(2)$ subgroup in the  Euclidian   $SO(5,1)$ $R$-symmetry, and then taking a diagonal of this $SU(2)$  with one  $SU(2)$ factor of the Lorenz $SO(4)$   symmetry  \cite{marcus}. When one compares the fields of the $\N=4$ theory in its four possible formulations (untwisted,   first  twisted,   second  twisted    or  third twisted formulation),   they   are  thus related by   linear mappings, which,  basically,  map spinor indices onto  Lorentz ones, using  the algebra of  Pauli matrices. We now show  directly how the third twist can be obtained form the first one.

 The first twist is that  we   used for writing the horizontality conditions in the precedent section, and that we justified by  straightforward dimensional reduction from  the 8-dimensional theory. It is
 \be
 \lambda 
\to ( \Psi_\mu,   \bar   \Psi_\mu, \chi_{\mu\nu^-},  \bar   \chi_{\mu\nu^-},    \eta,     \bar \eta   )
\ee
 \be
\phi\to (\Phi,\bar \Phi, L, h_{\mu\nu^-})
\ee
The 16 generators of the susy algebra which compose $\mQ$  and their    parameters
$\epsilon$ are     respectively  twisted into 
$Q_0,\bar Q_0,Q_\mu,\bar Q_\mu,      Q_{\mu\nu^-},\bar   Q_{\mu\nu^-}$ and   
\be
 \epsilon 
 \to ( \epsilon _0,     \bar  \epsilon _0 ,      \epsilon _\mu,   \bar    \epsilon _\mu,  \epsilon _{\mu\nu^-}, \bar    \epsilon _{\mu\nu^-} )
\ee
where we have identified      the index $I$ with a self dual index   $\mu\nu ^{-}$.  To suppress all equations of motion in $\mQ^2$, one  must have  a symmetry with not more than    9 generators  \cite{N=4}. To  define  the third twist from the first one,   we can heuristically 
 choose a restricted  7-dimensional family  of parameters,   made of the 4 scalars  $u , \bar u, v,\bar v$
 and a vector  $\kappa$,  where $\kappa^\mu$ is  of norm 1.  It reads
\be  \epsilon_0=u  \ \ \ \  \bar  \epsilon_0 =    \bar u  \ \ \ \  
  \epsilon _\mu =v  \kappa_ \mu,  \ \ \ \       \bar  \epsilon _\mu  = \bar  v   \kappa _\mu
\ee
 and   $ \epsilon _{\mu\nu^-}=   \bar    \epsilon _{\mu\nu^-}  =0$. One   has    a nilpotent      differential $  d+\mQ
 -    i_{(u \bar v  + v \bar u)    \kappa}   $.
The expression for the action of $\mQ$ on $A$ and $c$  is   obtained by decomposing the following graded equation in form degree
 \bea\label{HOR}
{\cal F} \equiv (d+\mQ
 - i_{(u \bar v  + v \bar u) \kappa } 
  )(A+c) +(A+c)^2=
F\CR
 + u \Psi + \bar u \bar\Psi 
  + \bar  v\scal { g(\kappa) \eta +i_ { \kappa} \chi_- }                                                      
 + v\scal{ g(\kappa) \bar  \eta +i_ { \kappa} \bar \chi_-}
 \CR                                                 
+(u^2+v^2) \Phi                                                                                       
+(\bar u^2+\bar  v^2) \bar\Phi                                                                     
+(u\bar u + v\bar v) L 
 \eea

The action of $Q$ on all fields in the right hand side of Eq.~({\ref{HOR}}) is obtained by expanding in form degree the Bianchi identity
$(  d+\mQ
 -    i_{(u \bar v  + v \bar u)    \kappa} ){\cal F}=-[A,{\cal F}]$.
 One has on all fields
\be
\mQ^2= \L_   {(u \bar v  + v \bar u) \kappa } 
\ee
 
 If we put $\bar u= \bar v=0$, we have  a  symmetry with  a family of  5 =2+3 parameters, namely    $u,v $ and $\kappa$. Then,      the restricted  differential operator $Q$  is nilpotent
\be
Q^2=0
\ee
 The $\mQ$-transformations are however still  $\kappa   $-dependent, according to
 \bea
 (d+\mQ)(A+c) +(A+c)^2=
F 
 + u \Psi                                                                                                                                   
 + v\scal{ g(\kappa) \bar  \eta +i_ { \kappa} \bar \chi_-}                                                  
+(u^2+v^2) \Phi                                                                                       
 \eea
Notice that that, by  this restriction on the parameters, the  fields     $\chi_- $  and $\eta$  have disappeared from   equations, so that their transformation laws are now   unconstrained by the curvature condition for $A+c$.

We now redefine    
\bea
\tilde \Psi \equiv   g(\kappa) \bar  \eta +i_ { \kappa} \bar \chi_-\eea
The troublesome dependence on  $\kappa$  can now        
    be forgotten,    by considering    $\tilde \Psi$ as an independent  new  1-form field,   $ (  \bar  \eta ,  \bar \chi_-)      \to  \tilde \Psi_\mu   $,      so that  the curvature condition  becomes  
  \bea \label{F_A}
 (d+\mQ)(A+c) +(A+c)^2=
F
 + u \Psi         + v\tilde \Psi                                             
+(u^2+v^2) \Phi                                                                                       
 \eea
Since it  now  depends only on scalar parameters, the   last   equation can   be extended       in curved space. $\mQ$  has become a true differential operator, which transforms fields into   expressions  depending  on both scalar real  parameters  $ u$  and $ v$. The intriguing fact, which we shall shortly exploit,  is the analogy     between  Eq.~(\ref{F_A})  and    the geometrical horizontality condition of the $\N=2$ theory (which is exactly reproduced for  $u=1,v =0$, as in \cite{basi}).

There is another  remarkable feature. If one looks  at the transformation laws of $h_-$ and $L$, as obtained from Eq.~(\ref{hch})  and  its  Bianchi identity   Eq.~(\ref{hcA}),  and if one  defines the following  one-form 
\be\label{V}
V \equiv   g(\kappa)  L +i_ { \kappa}  h_-
\ee
we  get another $\kappa$-independent equation satisfied by $V$, as follows 
 \bea\label{F_V}
 (d_A+\mQ)V +[  A+c,  V]=
d_A 
 -v  \Psi        +u      \tilde \Psi                                                                                                                               
 \eea
By combining Eqs.~(\ref{F_A}) and ~(\ref{F_A}), we get the following  complex equation, where, for the moment,  $u$ and $v$, as well as all fields,  must be considered  as real 
 \bea \label{F_Ac}
 (d+\mQ)(A+iV+c) +(A+iV+c)^2=
F_{A+iV}
 + (u -iv)    (\Psi         + i\tilde \Psi   )                                          
+(u^2+v^2) \Phi    
 \eea
By analogy with Eq.~(\ref{V}), we redefine     the 1-form $ \bar \Psi$   as a    an  antiself-dual 2-form    $ \chi _{\mu\nu^+}$  and   a scalar      form $\chi$   as follows 
 \bea
   \chi    &\equiv& i_\kappa \bar \Psi
\CR
     \chi _{\mu\nu^+}     & \equiv&        \kappa_{[\mu}  \bar \Psi  _{\nu]^+}    \eea
     These changes of variables have transformed the fields 
     $ \Psi_\mu,   \bar   \Psi_\mu, \chi_{\mu\nu^-},\eta,  \bar   \chi_{\mu\nu^-},  \bar \eta ,
      \Phi,\bar \Phi, L, h_{\mu\nu^-}$ of the first twist into the set of  fields
       \be A_\mu, V_\mu,   \Psi_\mu,\tilde \Psi_\mu,  \chi_{\mu\nu^-},    \chi_{\mu\nu^+},    \chi,\eta, \bar \Phi,\Phi\ee
       which is   nothing but the set of fields that Marcus  found  in its  direct third twist of the $\N=4$ formulation. Here, this set has been found   by the requirement of  getting a  geometrically meaningful  curvature  equation  involving all   fields  of the  $\N=4$ theory.  
       
    This construction  is quite  suggestive. Moreover, the complex  horizontality condition~(\ref{F_Ac})  indicates  a   closer relationship  between the $N=2$ and $\N=4 $ theories. We will find that the $\N=4$ theory, expressed in its third twist formulation is a mere complexification of the $\N=2$ theory, associated to a suitable supersymmetric  gauge-fixing of its imaginary part.

     \section{Complexification of the $\N=2$  Yang--Mills supersymmetry}
 
 \def\Lim{ {\cal L  }_{\rm Im}}

\subsection{Complex gauge invariance with twisted $\N=2$ supersymmetry}
\def\t{\tilde}
At this point, it is best to reconsider the problem, and come to the point of this presentation, by starting with   a complex  gauge field $A+iV$, within the context of $\N=2$ supersymmetry. We   therefore enter the not so familiar  domain of a gauge theory where   the infinitesimal parameter is complexified
$\epsilon \to  \epsilon +i\t \epsilon $. 
\def\O{\Omega}
The Fadeev-Popov ghost is then complexified,  $\O  \to  \O+i \t \O$,  as well as the antighost and lagrange multiplier field, and one  has  the  ``complex"   BRST symmetry
\bea
s (A+iV)  &= &-d  (\O+i\t \O)-[    A+iV,  \O+i\t \O] \CR
s (\O+i\t \O)  &= &-   \frac{1}{2}  [  \O  +i\t \O,  \O+i\t \O] 
\eea
The symmetry equations have real and imaginary parts that must be considered independently, so that
$sA=-d_A\Omega +[\t\O,V]$  and   $sV=-d_A\t \O -[\O,V]$. The complex curvature has real part
$dA+AA -VV$ and imaginary part   $d_AV$.
For $\t \O =0$, one has the ordinary   ``real"  BRST symmetry, for which $V$ only transforms  tensorially.

A complex gauge theory is problematic at the classical level. Indeed the Yang--Mills term 
   ${F_{A+iV}}  ^*  F_{A+iV} $ is invariant under the complexified symmetry, but   it is complex. In this action, both quadratic    parts  for $A$ and $V$ are  however transverse, and need gauge-fixing.  On the other hand   
 the real action ${F_{A+iV}  }^*  F_{A-iV} $    is only $s$-invariant when   $   \t \O=0$. This action has   purely   transverse propagators too, both for $A$ and $V$.
 We will shortly  see that, within the context of the $\N=2$ supersymmetrization of the complex  gauge theory,    the condition  $\t \O=0$   can be obtained  from a gauge-fixing that eliminates some of the fields, in a supersymmetric way.
 
\def   \ik{i_\kappa}
 \def   \gk{g_{\kappa}}
 \def\Qv{Q_\kappa}
 \def\c{{\gamma_1}}
As in the case of  the standard ${\N}=2$ supersymmetric theory,   one may try to   construct   five generators made of a  scalar  and   a vector  differential operator    $Q$ and  $Q_\mu$,  with     $Q_\kappa \equiv  \kappa^\mu  Q_\mu$.  
Here   $\kappa$ is a  commuting vector field  with   4 real components
  $\kappa^\mu$. $Q$ is made of two independent  scalar generators $Q_1$ and $Q_2$, with
$Q=uQ_1+vQ_2$, where $u$ and $v$ are two independent parameters. 
Eventually,  these  parameters should be interpreted  as  those that  one can conventionally  obtain by twisting the Poincar\'e $\N=4$ supersymmetry. But, here, their  introduction is motivated from a another point of view.  
Complexified   scalar and vector   shadows  must be defined,
\be    c\to c+i\t c    \ \ \ \      \c_\mu dx^\mu   \to     ( \c_\mu  +i\t \c _\mu)  dx^\mu
\ee
One has the definition
\def\A{\mathcal A}
\bea
\A
\equiv 
A+iV  +  
( u+iv )
 ( c+i\t c    )
+\ik(\c  +i\t \c ) +   \O+i\t \O 
\eea
where $\O+i\t \O  $ is the complex  Faddeev--Popov ghost.

 \def\z{u+iv}
\def\zb{u-iv} 
\def\zm{u^2+v^2}

The goal is is of  building  $Q$ and  $Q_\mu$,  with \bea
Q^2= Q_\mu Q_\nu +Q_\nu Q_\mu=0\ \ \ \ \ \  Q Q_\mu +Q_\mu Q\sim  \partial _\mu
\eea
on all complexified fields.
Here will reduce ourselves to the definition of the scalar operator $Q$, whose action involves the two parameters $u$ and $v$. 

Thus, from now on we set $\kappa=0$, which amounts to eliminate  the dependence in      the vector ghosts, and we have the restricted definition
\def\A{\mathcal A}
\bea
\A
\equiv 
A+iV  +  
( \z )
 ( c+i\t c    )
+   \O+i\t \O 
\eea

The  matter  fields of the   twisted $\N=2$ theory are extended as follows
\be
\Psi \to  
\Psi  +i  \t \Psi \ \ \ \
\Phi \to  
\Phi  +i  \t \Phi \ \ \ \
\bar \Phi   \to    \bar \Phi  +i  \t   {\bar \Phi}
\ \ \ \     \eta  \to  \eta  +i \t   \eta 
\ \ \ \       \chi_-    \to  \chi_-  +i \chi_+
   \ee
The standard horizontality condition \cite{basi}  of the $\N=2  $   theory is ``naturally"  complexified as follows
\def\zz{(u-iv)  }
\def\zzb{(u+iv) }
\bea
\big(  \ d+s+Q
 \big)  \A
+\frac{1}{2}   [\A,\A]    =
F_{A+iV} 
+(u-iv)
(\Psi +i  \t \Psi)
+
(u^2+v^2)  (\Phi  +i  \t \Phi )
\eea
with the Bianchi identity
  \bea
\big( \ d+s+Q
  +[\A,\ \ ]    \ \big)    \Big ( 
F_{A+iV} 
+(u-iv)
(\Psi +i  \t \Psi)
+
(u^2+v^2)  (
\Phi  +i  \t \Phi) 
 \Big)  =0\CR
\eea
These equations  are completely  analogous to those of   the ordinary case of the $\N=2$ theory. They determine the action of  both  $s$  and 
$Q$,  by expansion in form degree, so that nilpotency is warranted  because of the Bianchi identities for all values of $u$ and $v$.   In this complexified case one has to further split the equations in their real and imaginary parts (assuming  for instance that the Lie algebra matrices are real, as well as $u$ and $v$). This yields the differential operator $Q$ whose  action on the classical fields $A, V,  \Psi$ can be identified  as the sum of two scalar  supersymmetries with parameters $u$ and $v$ and   gauge transformations with   parameters  equal to the shadows   $c$ and $\tilde c$.

 


Following the   method of     \cite{shadow},   one     introduces  the BRST-partners of the shadows, to make the later ones  parts of a BRST-exact doublets. This permits one 
  to determine  the action of the BRST symmetry operator $s$  on the shadows and that of  $Q$   on the ghosts.  One    has   also     a  complex     BRST-exact doublet for the    anti-ghost  sector, which is  made of $  \bar \O      +
i      \t{\bar \O    }      $     and    $ H+i \t { H}  $, with
\bea
s ( \bar \O      +
i      \t{\bar \O    } )
=  H+i \t { H} 
-[  \O+i\t \O,   
 \bar \O      +
i      \t{\bar \O}	 ]
 \ \ \ \    s(  H+i \t { H})  =-[    \O+i\t \O,   H+i \t { H}]
\eea
The   BRST exact doublets in the shadow sector are  
$\bar c+i \t {\bar c}$,
$  \mu+i \t {  \mu }$ and  $\bar \mu+i \t {\bar \mu }$, with
\bea
s (c+i\t c)
=  \mu+i \t {  \mu },  \ \ \ \    s(  \mu+i \t {  \mu })  =0\CR
s(\bar \mu+i \t {\bar \mu })   
=\bar c+i \t {\bar c}    \ \ \ \    s(\bar c+i \t {\bar c})   =0\
\eea
\def\Qc {Q_{\hat c} }
\def\Pt{ \Phi+i\t \Phi}
\def\Pt{    (u^2+v^2)( \Phi+i\t \Phi)}
\def\re{    (u^2+v^2) }

It is useful to list the following 
 $Q$-transformations of the fields, as given by
 the horizontality equations.
 \def\ct{c+i\t c}
  \bea
 Q  (A +iV) &=&  (u-  iv) (\Psi+i  \t  \Psi)        -\zz d _{ A +iV}    (\ct)   \CR
Q  (c+i\t c   ) &=&\zzb   (  \Phi+i\t \Phi  )   - \zz   (c+i\t c) ^2       \CR
 Q  (\Psi+i  \t  \Psi)   &=&  (u+iv) d _{ A +iV} ( \Phi+i\t \Phi)  -\zz  [ \ct ,   \Psi+i  \t  \Psi  ]  \CR
  Q  (\Phi+i  \t  \Phi)   &=&    - \zz  [ \ct ,  \Phi+i  \t  \Phi ]  
 \eea
 \bea
\Qc (    \O   +i\t \O  )
=  -\zz  (    \mu  +i   \t {  \mu } )              \ \ \ \     \ \ \ \    \Qc  (  \mu+i \t {  \mu })  = 
-    \zzb  
[ \Phi  ,    \O   +i\t \O ]
\CR
\Qc (   \bar  \mu+i   \t { \bar    \mu }    )
=   \zz (  \bar \O      +
i      \t{\bar \O    }     )       \ \ \ \     \ \ \ \    \Qc  (  \bar \O      +
i      \t{\bar \O    } )  =\zzb    
[ \Phi,   \bar  \mu+i   \t { \bar    \mu }     ]
\CR
\Qc (   \bar c   +i   \t { \bar   c}    )
=   -  \zz (H       + i      \t{H }       )      \ \ \ \      \ \ \ \      \Qc  (   H      +  i      \t{H    } )  =-\zzb 
[ \Phi,     \bar c+i   \t { \bar   c}  ] 
\eea
where $\Qc\equiv   Q       -\zz [c+i \t c,\ \ ]$, and thus 
$\Qc^2 =(u^2+v^2)[\Phi +i \t \Phi,  \  ]$.
 These equations must be decomposed in real and imaginary parts, which yields the action of $Q_1$ and $Q_2$ on all fields from the $u$ and $v$ dependence of the $Q$ transformations. 



In view of a the further determination of a $Q$ invariant action, we need two introduce
pairs of  complex     self-dual and antiself-dual   anticommuting 2-forms $\chi_-$ and $\chi_+$ with      Lagrange multipliers  $H _\pm$,
as well a complex  commuting scalar fields  $ \bar   \Phi+i\t {\bar \Phi }$  with fermionic lagrange multipliers    $  \eta+i     \t {  \eta }$, 
 such that  
 
  \bea
\Qc (   \bar   \Phi+i\t {\bar \Phi } )
= \zz       (  \eta+i     \t {  \eta }    )    \ \ \ \      \ \ \ \      \Qc (   \eta+i     \t {  \eta }   )  = 
    \zzb   [ \Phi,    \bar   \Phi+i\t {\bar \Phi }         ]
  \CR
 \def  \Qcc{{Q_{c+i\t c}}}    
\Qc  \chi_  \pm  =  \zz  H_\pm    \ \ \ \      \ \ \ \ 
\Qc    H_ \pm      =      \zzb     [\Phi +i \t \Phi,    \chi_  \pm ]
\eea 
The introduction of the    fields   $ \chi_  \pm$ and     $    \bar  \Phi+i\t {\bar \Phi }$  would be more natural within the context     of  the vector symmetry transformations that we donnot discuss here \footnote{There is no reality condition on $\chi_+$ and $\chi_-$, as well as on $H_+$ and $H_-$,  but each one of these fields  counts for 3 degrees of freedom, as a self-dual or antiself-dual field. }.

   .

\section{  Equivariant gauge-fixing with complexified self-duality equation for $A+iV$}


The full     gauge symmetry includes  imaginary and real parts, and we  might    look for  the      obtention of a  $Q$-invariant action in  the cohomology of the complex BRST symmetry.  However, by doing, and using a  a $Q$-exact action,     one   gets   an  action  of the type 
$\int  (  F_{A+iV}  \wedge ^*  F_{A+iV} +{  \rm  supersymmetric  \  terms})$, which is interesting per se, but doesn't reach our      goal  of getting a unitary theory such as the $\N=4$ theory.  Rather, we  will   look for  the      obtention of a  $Q$-invariant action in   the cohomology of the real part of the  BRST symmetry.   This means that must take the  condition $\t c=0$ for  the definition of $Q$ as well as $\t \O=0$ for that of $s$.  In the next section, we will show that the restriction of the BRST symmetry to its real part can be done by a gauge-fixing that uses a $Q$-invariant action.

One        wishes  a     Q-exact action whose $u,v$ dependence is only an overall factor of $u^2+v^2$, modulo boundary terms that can depend on   $u$   and $v$, at least after the elimination of auxiliary fields, as in \cite{KW}. This ensures the  invariance of the  action  under both  $Q_1$ and $ Q_2$.

The   following action,   which   turns out to use    complex self-dual conditions  as  a $Q$-antecedent,  satisfies such    requests :
\begin{multline}
I= Re \   Q \int \trace    \Big(     (  \chi _-  -i    \chi  _+  )  \wedge  (    {   \cal F  +\star {\bar{ F} } }-      \frac{u-iv}{2} ( H  _-+iH_+  ) )  
 \Big )\\*
=  Q \int \trace \Big  (       \chi_ - \star \scal{u ( F_A -VV )  -v  d_A V )_-   -  \frac{u-iv}{2}  H_-      }\\*
+      \chi_ + \star \scal{v  (  F_A -VV ) +u   d_A V )_+  - \frac{u-iv}{2}  H_+   }
 \Big )\\*
=   \int  \trace \Bigl(- \frac{1}{2} | H|^2 
+        H_-\star \scal{u ( F_A -VV )  -v  d_A V }_-
+ H_+\star \scal{v( F_A -VV )    +u d_A V }_ +\\*
 (u^2+v^2)
 \scal{ \chi_- \star( d_A   \Psi-   [V,\tilde \Psi] )  +    \chi_+ \star (d_A \tilde \Psi    +[V,  \Psi])
  -\Phi    | \chi|^2  } \Bigr)
\end{multline}
where  $  | \chi|^2    = \frac{1}{2} [\chi_+, \star \chi_+] +  \frac{1}{2}[\chi_-, \star \chi_-]$.  We have redefined $H_\pm$ in such  way that 
$Q  \chi _\pm  =(u-iv)H_\pm,   QH_\pm=0$.

 We have introduced
\be { \cal F} \equiv(u+iv)   F_{A+iv}\ee and    one has    \be{   Q   \cal F} =(u^2 +v^2)    (     d _{A+iV}
(\Psi+i\t \Psi)   -[c+i\t c,   { \cal F}  ])
\ee
with   $s   { \cal F} =      -[\O  +i\t \O  , { \cal F}  ]$.

When one integrates over $H_\pm$,   $(u^2+v^2)$ factorizes and one gets, 
\begin{multline}\label{lagI}
I 
=    (u^2+v^2)\int \trace \Bigl(
 \frac{1}{4}   (  F_A -VV   +i  d_A V ) \star \overline{( F_A -VV    +i d_A V )}
\\*
 +
 \chi_- \star ( d_A   \Psi-   [V,\tilde \Psi] )   +    \chi_+ \star  (d_A \tilde \Psi    +[V,  \Psi])  -\Phi  | \chi|^2 
 \\*
-Re \scal{ (u+iv)^2  \frac{1}{4} (  F_A -VV   +i  d_A V )_{\, \wedge }(  F_A -VV   +i  d_A V )} \Bigr)
\end{multline}
 The desired result  is present : there is not $u,v$-dependence in  $I  /   (u^2+v^2)  $, but for the last term  that   is a topological term for the complex connection $A+iV$
 \be
 I_{\rm top}=\int
   \frac{1}{4} Re \scal{ (u+iv)^2 \trace  (  F_A -VV   +i  d_A V )_{\, \wedge }(  F_A -VV   +i  d_A V )}
 \ee
  
 The latter  topological term is defined in function of     
  \be \trace \F\wedge \F = (u+iv)^2 
\trace F  _{A+iV}\   \wedge F  _{A+iV}  =    (u+iv)^2  
\trace  ( F_A \wedge F_A    + 2i   \trace  d (V  \wedge  F_A))       \ee 
 where  $ V  \wedge  F_A$ is globally well defined when one restricts the gauge invariance  to its real part.   It 
  can be thought of as a classical lagrangian, (in the spirit of TQFT's as in \cite{basi}),  such that  one has locally
\bea
\trace  (  F_A -VV   +i  d_A V )      \wedge (  F_A -VV   +i  d_A V )    \CR =
 d \trace   (       ( A+iV)  F_{A+iV}-\frac{1}{3}  ( A+iV) ^3   )
 = 
 d\trace    (A F_A-\frac{1}{3}  A^3   +2i  VF_A)
  \eea  
 The       action   (\ref   {lagI})  is its gauge-fixing, using self-duality gauge conditions, analogously as in~\cite{basi}.
 
 So,    the   ``topological gauge functions"     for $A+iV$  are 
  \bea
    (    u (F_A -VV )  -       v    d_A V) _-  \CR
    (v   ( F_A -VV) +u     d_A V )_+
   \eea
These gauge conditions have   the  following suggestive  expression    in complex notation 
   \bea\label{sd}
   (u+iv ) (  F_A -VV   +i  d_A V )  -   \star {\overline  {(u+iv )  (  F_A -VV   +id_A V )  }}=0
   \eea
   that is
   \bea
       { \cal F  =  \star {\bar{ F} }} 
       \eea
   which    expresses a class of self-duality  condition of $A+iV$, parametrized by the parameter $u/v$, and covariant under the real part of the gauge symmetry. The action is thus obtainable by a generalization of the method of \cite{basi}, which was well-suited for the Donaldson-Witten theory, but here we have used a complexification that amounts to a doubling of all fields.   
   
   At this point we have the following observations.


  The six components of the self-duality equations are such that  $I$ gives transverse propagators,  both for $A$ and $V$.  Their   gauge-fixing  can be possibly done in a rather standard way, by choosing a Landau gauge for both $A$ and $V$, which  gives a theory that  has yet no interpretation. In particular, it is not the $\N=4$ theory. Moreover, at this stage, there is a gauge degeneracy for the propagators of  $\Psi$ and $\t \Psi$, and the propagation of the field $\Phi$  has not been    ensured.   If we count the number of on-shell degrees of freedom of the fields $V_\mu, \Phi, \bar\Phi,  \t\Phi, \bar{\t \Phi}$, we find however 2+1+1+1+1=6 degrees of freedom, which is  also  the number of scalar fields of the $\N=4$ theory.
  
  We now come to the point of gauge-fixing the    imaginary part of the gauge symmetry for recovering the $\N=4$ theory in its  third twist formulation, and ensuring a standard propagation for all fields. 
   

%

 \section{ $Q$-invariant gauge-fixing  for the imaginary part of the gauge symmetry}
 
 \subsection{Elimination of the imaginary part of   ghosts and shadows}
 The following quantities with  shadow number -1   are invariant under the  BRST  symmetry   of  the real part of the gauge symmetry (with $D\equiv d_A$)
  \bea
 \t      {\bar  \Phi }   \t    c   \ \ \ \ \ \    
      \t{\bar \mu   } \t \O        \ \ \ \ \ \   
 (a   \t{\eta  }+b     \t{\bar c  })(  D^\mu V_\mu +  c{\t H}   )    \ \ \ \ \ \   
      \t{\bar \Phi  } (  \alpha    D^\mu   \Psi _\mu    +  \beta       D^\mu  {\t  \Psi }_\mu   -\delta  [\Phi, \eta]) \CR
  \eea
  where $a,b,c, \alpha,\beta,\gamma, \delta $ are numbers.
  The   $Q$-invariant term 
  \def\Lim{ {\cal L  }_{\rm Im}}
\bea\label{gfi}
\Lim=
Q(\t  
 {\bar  \Phi }     \t    c
 + 
      \t{\bar \mu   } \t \O)
  )\CR
      = \t    {\bar  \Phi }  
       \t \Phi     + \t \eta      \t    c    
       +
         \t{\bar \mu   } \t \mu      +  \t{\bar {\O}}     \t \O      
            \eea           
 breaks the imaginary part of the gauge symmetry, but respects its real part. It  yields  algebraic  equations of motion of  action   that enforce the condition
  \bea\label{cond}
 \t  \O=   \t{\bar {\O}}  
 = \t    {\bar  \Phi } 
 = \t      \Phi =
 \t  \eta=0
 \eea
for having only the  real  part of the gauge symmetry. 
 
After this elimination,  the following supersymmetric term 
\bea\label{gfii}
    Q\Big (        \t{\bar c }   (D^\mu V_\mu +      {\t H}   )
       +       \t{\bar \Phi  } (  \alpha    D^\mu   \Psi _\mu    +  \beta       D^\mu  {\t  \Psi }_\mu   +\delta  [\Phi, \eta]) 
   \Big )     
  \eea 
yields an action 
 with the following    form  
\bea
 \t H^2  +\t H     D^\mu V_\mu   
 +
    \eta   D^\mu \Psi  _\mu     +    \t{\bar c }       D^\mu\t  \Psi  _\mu   +\bar \Phi D^2  \Phi+\ldots
\eea
  It defines a    longitudinal propagation of $V$   through a   term
   $ \sim  | D^\mu V_\mu |^2   $, as  well as a    longitudinal propagation   for   $  \Psi $ and   $\t  \Psi$ by the terms
  $ D^\mu V_\mu  $ and  $D^\mu \tilde\Psi_\mu$.  There is some flexibility for the relative coefficients if one only requires $Q$ and BRST invariance. However, the demand of vector symmetry for the sum of both   action~(\ref{gfii})  and~(\ref{lagI}) fixes all the coefficients.
  
  In fact  $ \t{\bar c } $ and 
  $\eta$ play the role of propagating fermionic Lagrange multipliers, and 
  $\Lim$ is an action that enforces the equivariant (with respect  to the real part of the gauge symmetry) topological gauge functions 
  \be\label{gfV}
D^\mu V_\mu +...=0
\ \ \ \ \ \ \ \ \ \ 
D^\mu \Psi_\mu +...=0
\ \ \ \ \ \ \ \ \ \ 
D^\mu \tilde\Psi_\mu +...=0
\ee
We have  the following   $Q$-quartet diagram
\bea
\begin{matrix}
&\ &  \bar   \Phi & \   \\
&\eta &  \ & \t {\bar c}  \\
&\ &  \t H & \   \\
\end{matrix}
\eea
These fields   transform  tensorially under the  $s$ transformations, for $\t \O=0$. The way  the anti-shadow field  $ \t {\bar c}  $    becomes associated to the field $\eta$ after the gauge-fixing of the imaginary part of the gauge symmetry     is quite interesting.

The property $Q_c^2 =(u^2+v^2)[ \Phi,\  ]$ can be  enforced as 
 \bea
 Q_c  (\eta  +i{\t    {\bar c}})  =(u-iv)   (   [\Phi,\bar \Phi]  +i   \t  H  )
 \CR
 Q_c    (  [\Phi,\bar \Phi]  +i   \t   H  )=
 (u+  iv)      [\Phi,   \eta  +i{\t    {\bar c}} ]
 \eea
 provided one does field rescalings. The first equation gives $Q{\t    {\bar c}}=u  H+...$, and the second one gives   $Q \bar \Phi =  u \eta+...$. 
The conjugate  operator  $  \bar Q$ that anticommutes  with  $Q$   is defined by
\bea
\bar Q_c  (\eta  +i{\t    {\bar c}})  =  { i} (u-iv)    ( [\Phi,\bar \Phi]  + i  \t   H  ] )
 \CR
 \bar Q_c    (  [\Phi,\bar \Phi]  +i\t  H  )=
- { i}  (u+  iv)     [\Phi,   \eta  +i{\t    {\bar c}} ]
 \eea
These equations determine the $Q$ and $\bar Q$ transformations of the  fields of the quartet
 $\bar \Phi, \eta, {\t    {\bar c}}, H$, by separation of their real and imaginary parts.

%
%
%


\section{Recovering the $\N=4$ theory}

\subsection{Restricted horizontality equation and twisted $\N =4$}
\def\A{\mathcal A}
After the $Q$-invariant gauge-fixing of its imaginary part, the BRST symmetry is 
\bea
s (A+iV)  &= &-d  \O -[    A+iV,  \O ] \CR
s (\O )  &= &-   \frac{1}{2}  [  \O ,\O] 
\eea 
It identifies $V$ as a vector that only transform tensorially.  By using the reduced  unified field 
$
\A
= 
A+iV  +   (u+iv)  c     
+\ik  \c   +   \O  
$, we have obtained 
  the following  equation for the  definition of $Q$ 
\bea\label{hhc}
(d+s+Q+
)\A
+\frac{1}{2}   [\A,\A]
= 
F_{A+iV} 
+(u-iv)
(\Psi +i  \t \Psi)
+(u^2+v^2)  \Phi   
\eea
  with  its Bianchi identity
  \bea
(d+s+Q
  +[\A,\ \ ])
 ( 
F_{A+iV} 
+(u-iv)
(\Psi +i  \t \Psi)
+
(u^2+v^2)\Phi   
)=0
\eea
%

These equations give  the two  scalar  transformation   laws  the $\N=4$ theory in the third twist,  
using   the field $A+iV$. 


 The    invariant action can be expressed in a most simple form, as the sum \begin{multline} \label{mp}
(u^2
+v^2)  I_T=    \int   Q \trace \Big  (        {\t    {\chi      }}_ - \star \scal{u ( F_A -VV )  -v  d_A V )_-   -  \frac{1}{2}  H_-      }\\*
+      {\t    {\chi  }}_ + \star \scal{v  (  F_A -VV ) +u   d_A V )_+  - \frac{1}{2}  H_+   } \Big )
\\*+
 \int    Q   \bar Q       \trace
 \Big  (        \eta  \star  {\t    {\bar c}}  +\bar 
\Phi   d_A\star V   \Big  )  
\end{multline}
The   $  Q   \bar Q $ exact term  reproduces the $Q$ invariant actions discussed in the last section for eliminating the imaginary part of   ghosts and shadows and providing the longitudinal part of $V$.
One can check that the action~(\ref{mp}) reproduces the  one originally found by Marcus for the $\N=4$ super-Yang--Mills action in the third twist   \cite{marcus}. 

The last term   of the action suggests the relevance  in perturbative theory of the    complex gauge function\footnote{The gauge -fixing of $A$ can be done by a $s$ exact term involving $H,\Omega$ and   $\bar\Omega$. Moreover, it can also be made $Q$-exact using the shadow fields as in \cite{shadow}. }
 \be
        d_{A  }   (A+iV)_\mu  =  \partial _\mu    A_\mu      +i          d_A   V  _\mu
\ee
 
What we have done is the following. 
     The complex self-duality equations count for 6~conditions.  A  seventh gauge condition  was needed   for producing a gauge-invariant    longitudinal term   of the form  $|d_A\star V |^2$,  since the   $\N=2$     action with complex  gauge symmetry (\ref{lagI})  only defines a transverse   propagation of $V$.  The last term in  $ I_T$  provides such a  term, as well as other needed terms for  defining  the propagation    of   the longitudinal  parts of $\Psi_\mu$ and $\Psi_\mu$ and of  all scalar fields of the complexified $\N=2$ theory. 
       We can reformulate the whole process by saying   that all the needed fields come from at complexified version of the $\N=2$ theory, followed by a supersymmetric gauge-fixing of the imaginary part of the gauge symmetry.  One has a sort of  transmutation between the imaginary parts of the  scalar fields and the  longitudinal degrees of freedom for the vector field~$V$.

        \subsection{ Supersymmetric observables   }

 One can  now simply observe  that,    with the condition  $u-iv=0$    on     the analytically continued  parameters  $u$ and $v$,  all gauge-invariant observables  ${\cal O }(A,F_A)$   determine   $Q$-invariant  quantities,  that are nothing but   $   {\cal O }  (  A+iV,    F_A -VV +id_A V)$.   Indeed,  for $u-iv=0$,  the  extended  curvature  condition~(\ref{hhc})  becomes identical  to that  for the   ordinary gauge invariance  
\bea 
 (d+Q+s )(A  +  i V   +c+\O ) +(A+iV +c+\O )^2= 
F_A -VV +id_A V    
        \eea
  Having such a  rich ensemble of  supersymmetric observables   has no equivalent in the twisted $N=2$ theory.  Kapustin and Witten discussed the Wilson and t'Hooft  loops for the field $A+iV$  as supersymmetric  observables.

        Otherwise,  for arbitrary values of $u,v $, the descent equations for the invariant polynomials of the  complex curvature hold and  give   $Q$-cocycles depending on $\Psi$ and $\Phi$, which  satisfy the   usual  relations   of  TQFT    ``topological observables", as in  \cite{basi}\cite{cswitten}.

We   may notice here    that   the   close  relationship between the derivation  both   twisted $\N=2$ and    $\N=4$  actions  suggests a     further relevance of    the  3-dimensional theory with  a  ``complex" Chern--Simons action  as in
\cite {cswitten}
\bea
\int_{\it M_3}
\trace   \scal{  ( A+iV)   F_{A+iV}-\frac{1}{6}  ( A+iV) ^3   
}  =\int_{\it M_3}
\trace    \scal{ (AF_A-\frac{1}{6}  A^3 )
+iV F_A
-Vd_AV
+
\frac{i}{6}  V^3
 }\CR
\eea
Natural observables  of this 3-dimensional theory are    Wilson loops of $A+iV$. One may question whether the  twisted  $N=4$ theory can be    derived from such a Chern--Simons theory, by   constructing a supersymmetric Hamiltonian 
$H\sim [Q,\bar Q]$ in three  dimensions, and extending it in four  dimensions, as a generalization of the method used by   Witten   for originally    constructing the topological twisted $\N=2$ action that describes  the  Donaldson invariants~\cite{DW}.

        \section {Topological sigma-model  and $2D$-gravity}
       
   Analogous extensions that use the complexification  of gauge symmetries  can be  done, for cases where horizontality conditions exist, such as the topological sigma-model and $2D$-gravity \cite{T2GSM}.         We will display formula that illustrate the method, and will remain at a very formal level.  We have chosen the most possible simple choices    of using the imaginary part of the gauge symmetries,  but more clever gauge choices might  exist.

       \subsection {Topological sigma-model}
   Let us first consider      the topological sigma model. Given a  world-sheet scalar  field $X^\mu$, in an appropriate    target space  with Kahler form  $J$, $J^2=-1$, its ordinary topological symmetry is \cite{T2GSM}
           \bea
       (d+Q )X=  dX +\Psi
       \eea
       The topological gauge function (holomorphic maps) is
            \bea
      \partial  X^\mu  =  J^\mu_\nu \bar  \partial  X^\nu
       \eea
       and the action is
       \be
       I\sim \int  d^2x \  Q \Big    (      \bar \Psi _\mu  (H^\mu +    \partial  X^\mu  - J^\mu_\nu \bar  \partial  X^\nu ) \Big)
       \ee
        
        One  can introduce a new scalar $\tilde   X  ^\mu$, and extend $X\to   X+i\tilde   X $.
        Then one  can generalize  $Q$ as  a symmetry with two parameters  $u$ and $v$, with
          \bea
       (d+Q )(X+i\bar X  )=  dX +id\tilde X+(u-iv) (\Psi  +i\tilde \Psi)
       \eea       One thus    has  a topological sigma-model, with  topological gauge function
         \bea
          \partial  (X+i\bar X)^\mu  =  J^\mu_\nu \bar  \partial  (X+i\bar X)^\nu
       \eea
       $Q$ has  2 generators, and is governed by the 2 parameters $u$ and $v$.   One  has descent equations, and the ordinary observables of the topological $\sigma$-model. However, for the values 
       $u=iv$, the correlators    of   $dX+i  d  \tilde    X$ are $Q$-invariant.
        \subsection {  $2D$-gravity}
        \def \z {{\bar z}}
        In topological gravity, the field is the  Beltrami  differential $\mu^\z_z$, its shadow is the anticommuting vector $c^z$, and we define
        \be
        \mu^z=dz+ \mu^z_\z   d\z 
        \ee
       Its topological symmetry
        involves the topological ghosts    $\Psi ^z_\z$  with
          \be
      \Psi ^z= \Psi ^z_\z  d\z
        \ee
        and ghost of ghost $\Phi^z$.
        The ordinary topological BRST operator $Q$ is given by 
         \bea
       (d+Q )(\mu^z  +c^z)   +(\mu^z  +c^z)    \partial _z  (\mu^z  +c^z)     =     \Psi ^z  +  \Phi^z
       \eea
        One  extends
        \bea
           \mu^z=dz+ \mu^z_\z   d\z  \to   dz+ (\mu^z_\z     +i V^z_\z   )d\z
        \eea
       and, as a generalization of the Yang--Mills case,  $ c^z \to     c^z  +i   {\tilde c}^z$.  Here we will chose $\t c ^z =0$. Then, one  redefines $Q$ into
         \bea
       (d+Q )(\mu^z  +iV^z+c^z)   +(\mu^z +iV^z +c^z)    \partial _z  (\mu^z +iV^z +c^z)     = \CR=          (u-iv) (\Psi ^z +i\tilde \Psi^z)  +   (u^2+v^2) \Phi^z\
        \eea
        so that  
        \bea  
        Q   \mu^z_\z   =  \partial_\z c^z  +c^z\partial _z       \mu^z_\z     -    \mu^z_\z   \partial _z                  c^z 
        +u\Psi^z  -v \tilde \Psi^z 
        \CR
              Q  V^z_\z   = c^z\partial _z       V^z_\z     -    V^z_\z   \partial _z                  c^z +
        v\Psi^z    +u  \tilde \Psi^z \CR
        Qc ^z= c^z\partial _z  c^z  + (u^2+v^2) \Phi^z\CR
        Q_c(\Psi ^z +i\tilde \Psi^z)   =(u-iv) \partial _\z  \Phi^z                  
        \eea    
        
        A $Q$-exact action that gives a $u$- and $v$-independent  action is  
        \bea
       I={ 1\over{u^2+v^2}}
       \int  d^2z\  \Large (  Q(  \bar \Psi _{zz} \mu^z_\z  +\bar Q  (
       \bar \Phi_{zz}   V^z_\z))\  \Large)
       \eea

       The elimination of the auxiliary   fields sets the  fields   $\mu,V $  and  all fermionic ghosts equal to zero. The only remaining propagating fields are  
       $b_{zz}=Q \bar \Phi_{zz} $ and $c^z$, and the action is still  the ordinary topological action
         \bea
        I\sim    \int  d^2z\  \Large (    b_{zz}   \partial _\z   c^z
        +  \bar \Phi_{zz}    \partial _\z   \Phi^z )
        \eea
        
       One  can  define observables as    $Q$-invariant correlators     for $u-iv=0$,    which  can be expressed in function of $\mu^z_\z     +i V^z_\z $.

 \end{document}